\documentclass[12pt,preprint2]{aastex}
\shorttitle{Dust Properties in the FUV in Ophiuchus}
\shortauthors{Sujatha et al.}

\begin{document}
\title{Dust Properties in the FUV in Ophiuchus}
\author{N. V. Sujatha, P. Shalima, Jayant Murthy}
\email{sujaskm@yahoo.co.in, shalima.p@gmail.com, jmurthy@yahoo.com}
\affil{Indian Institute of Astrophysics, Koramangala, Bangalore - 560 034, India}
\and
\author{Richard Conn Henry}
\email{henry@jhu.edu}
\affil{Department of Physics and Astronomy, The Johns Hopkins University,\\ Baltimore, MD 21218}
                                                                                
\begin{abstract} 
We have derived the albedo ($a$) and phase function asymmetry factor ($g$)
of interstellar dust grains at 1100 \AA ~ using archival {\it Voyager} observations of diffuse
radiation in Ophiuchus. We have found that the grains
are highly forward scattering with $g =$ 0.55 $\pm$ 0.25 and $a =$ 0.40 $\pm$ 
0.10. Even though most of the gas in this direction is in the Ophiuchus 
molecular cloud, the diffuse FUV radiation is almost entirely due to scattering 
in a relatively thin foreground cloud. This suggests that one cannot assume 
that the UV background is directly correlated with the total amount of gas in 
any direction.
\end{abstract}

                                                                                
\section{INTRODUCTION}
The Ophiuchus molecular cloud is one of a large complex of clouds at a 
distance of about 160 pc from the Sun \citep{chen97}. The region has been 
extensively studied in CO \citep{geus,Dame} as well as with four-colour photometry
\citep{Co04} allowing us to determine the three-dimensional distribution of the 
matter in this direction. Our interest in this region is that one of the first 
observations of diffuse emission in the far ultraviolet (FUV -- 912 - 1216 \AA) was 
made here by \citet{Holberg} who identified the emission as starlight from 
the nearby Scorpius-Centaurus OB association scattered by dust in the Ophiuchus 
molecular cloud.

Because of the technical difficulties inherent in diffuse observations in the FUV, 
largely due to scattering from the intense geocoronal Ly$\alpha$ line, there have 
been very few observations of the background radiation in this spectral region  
\citep[see][for reviews of the diffuse radiation in both the near and far ultraviolet]
{Bowyer91, RCH91}. By far the largest and most reliable data set has come from the 
ultraviolet spectrographs (UVS) aboard the two Voyager spacecraft which made observations 
of various
astrophysical targets during the interplanetary phase of their mission, between their
hugely successful planetary encounters. Many of these targets were of objects with no
intrinsic FUV flux and \citet{Mu99} used them for a comprehensive study of the diffuse
FUV radiation field. Because of the sensitivity of the Voyager UVS to diffuse radiation
and the distance of the spacecraft far from the Earth where emission from interplanetary
H {\small I} is minimized, these remain the definitive observations of the diffuse 
radiation field in the FUV.

Except for a small extragalactic component at high latitudes \citep{RCH91}, the
diffuse UV radiation is largely due to starlight scattered by interstellar dust 
and as such can be used to derive the scattering properties of the interstellar 
dust grains. Unfortunately there has been considerable controversy about both 
the level of the diffuse radiation and the modeling used to extract the optical 
constants \citep[see][for a discussion of the difficulties]{Dr03, Mathis02} and 
there have been only loose observational constraints on the albedo ($a$) and 
phase function asymmetry factor ($g$) of the dust grains.

As mentioned above, \citet{Holberg} discovered intense diffuse emission from Ophiuchus 
using the Voyager UVS and interpreted this emission as starlight back-scattered by the
Ophiuchus molecular cloud. We have reexamined these observations along with others from 
the Voyager archives \citep{Mu99} in the light of a more sophisticated model and an 
improved understanding of the dust distribution in the direction of Ophiuchus 
\citep{Co04} and have found that the emission is actually due to the forward scattering 
of the light from a much lower density sheet of material in front of the Ophiuchus 
molecular cloud. There were many observations spread throughout the entire region and 
so  we have been able, for the first time, to remove the degeneracy between the albedo 
($a$) and the phase function asymmetry parameter ($g$) which have plagued studies of 
the diffuse radiation \citep{Dr03}. We have constrained $a$ to 0.40 $\pm$ 0.10 and $g$ 
to 0.55 $\pm$ 0.25, in good agreement with the theoretical prediction of \citet{Wei01} 
for a mixture of graphitic and silicate grains.

\section{OBSERVATIONS}

The two Voyager spacecraft include identical Wadsworth-mounted objective grating 
ultraviolet spectrographs (UVS) covering the wavelength region between 500 and 1700 \AA, 
with their greatest sensitivity at wavelengths below 1200 \AA. The field of view of the 
spectrographs is $0.1\degr \times 0.87\degr$ with a spectral resolution of 38 \AA\ for 
diffuse sources. The spacecraft were launched within two weeks of each other --- Voyager 
2 in 1977 August and Voyager 1 in 1977 September --- and obtained a wealth of information
 on all four of the giant planets. The two spacecraft are still continuing operation at 
the edge of the solar system with more than 10,000 days of operation each. A full 
description of the UVS spectrographs and further information about the interstellar 
mission of the Voyager spacecraft is given by \citet{Holberg92}.

\begin{figure}
\plotone{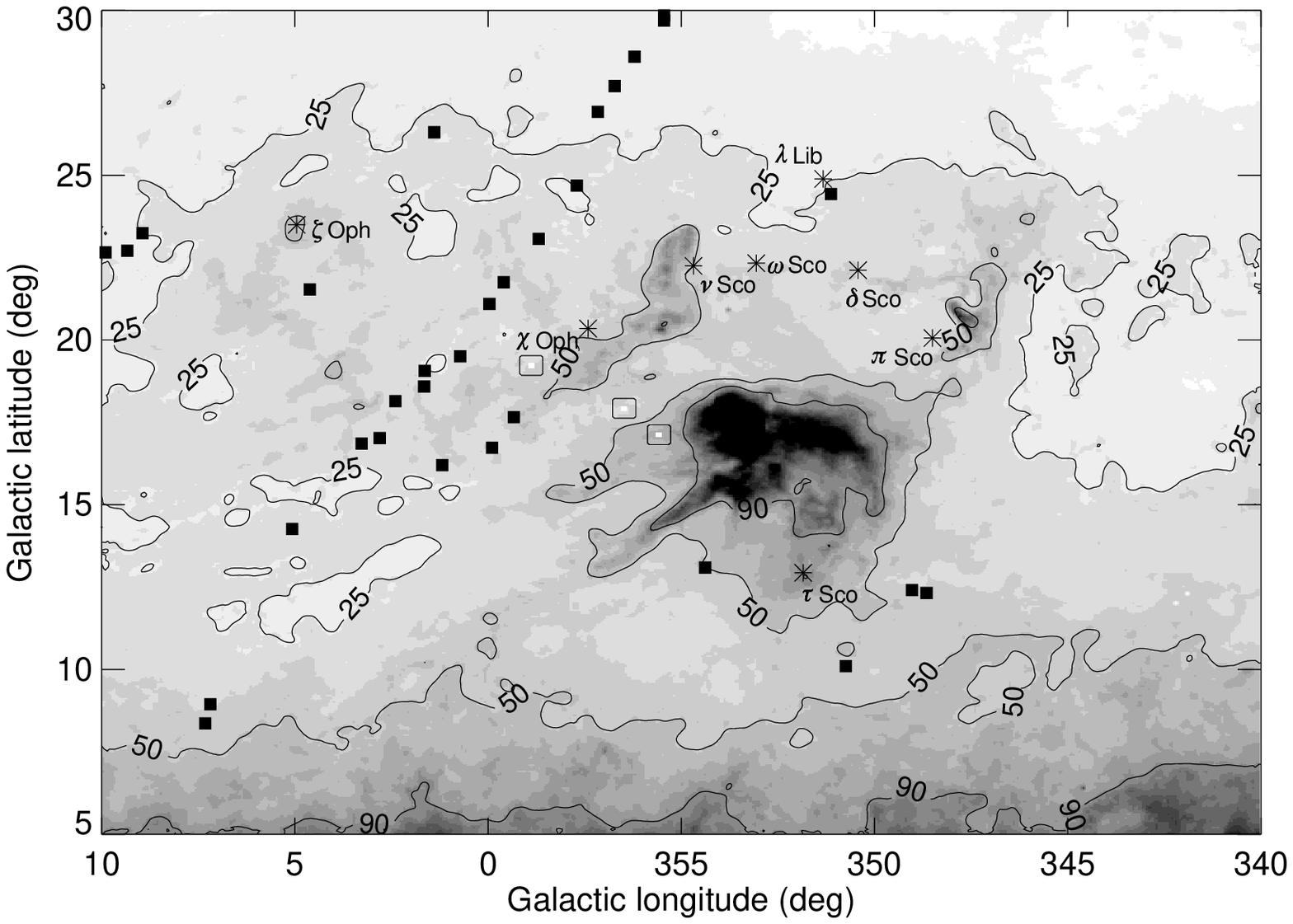}
\caption{IRAS 100 $\micron$ map of the region is shown with contours
labelled in units of MJy sr$^{-1}$. The filled squares show the locations of the 
{\it Voyager} observations and the asterisks show the positions of the brightest UV 
stars in the region.}
\label{f1}
\end{figure}

While the spacecraft were between planetary encounters, they observed many astronomical
targets. Amongst these targets were a series of scans in the vicinity of Ophiuchus
with the Voyager 2 spacecraft in 1982 \citep{Holberg}. We have supplemented these with 
further observations from the Voyager archives taken at various times between 1982 and 
1994 \citep{Mu99}. There were a total of 31 such locations and these are plotted on an 
IRAS 100 $\micron$ map in Fig. \ref{f1} and tabulated in Table \ref{t1} with the total 
hydrogen column density \citep{Sch98} and the IRAS 100 $\micron$ flux \citep{wheel}. 
There was a 
limit cycle motion of a few tenths of a degree in the Voyager pointing and the position 
reported in the table is an average of the actual pointing of the instrument. 

The data used here are from \citet{Mu99} and are available from the authors. The data
reduction is fully described in that paper and in \citet{Holberg86} and consists of 
fitting three components to the observed signal: dark noise from the spacecraft's 
radioisotope thermoelectric generator (RTG); emission lines from interplanetary 
H {\small I}; and emission from cosmic sources. The dark noise was subtracted using 
the continuum below the Lyman limit 912 \AA\ and the astronomical and heliospheric 
emission were then simultaneously fit using templates for the emission. This reduction 
procedure was shown to be consistent and reproducible over observations separated 
widely in time and between the two spacecraft and yielded 1$\sigma$ limits as low as 
30 photons cm$^{-2}$ s$^{-1}$ sr$^{-1}$ \AA$^{-1}$ over a large 
part of the sky. For the bright sources reported on here, the diffuse emission 
dominated the raw data and there was little uncertainty in the derived levels.

\section{MODEL}

The diffuse emission in Ophiuchus is the result of light from the nearby stars 
scattering from interstellar dust in the line of sight and were modeled in a
similar manner to \citet{Sh04} in the Coalsack. The stellar radiation field was 
calculated at the location of the scattering dust using the model of 
\citet{Suj04}, in which the distance and spectral type of each star was taken 
from Hipparcos data \citep{Perryman} and the flux calculated using Kurucz models 
\citep{Kurucz} with the latest modifications from his web page 
(http://kurucz.harvard.edu). Eight stars (see Fig. 1) contribute $\sim$90\% of the 
total ISRF in this region. The properties of the stars are given in Table \ref{t2}.

The amount of radiation scattered to the observer is dependent on the 
scattering function of the grains and we have used the Henyey-Greenstein 
scattering phase function, $\phi(\theta)$ \citep{HG}.
\begin{equation}
\phi(\theta) = \frac{a}{4\pi} \frac{(1-g^{2})}{(1+g^{2}-2gcos(\theta))^{1.5}}
\label{eq1}
\end{equation}
Here $a$ is the albedo --- which can range from 0 for dark grains to 1 
for fully reflecting grains --- and $g$ is the phase function asymmetry factor 
with $g = 0$ indicating isotropic scattering and higher values indicating 
forward scattering grains.

\citet{Draine} has suggested that the Henyey-Greenstein function 
underestimates the scattered radiation for highly forward scattering grains 
($g > 0.7$) in the FUV. We have found, empirically, that using the theoretical 
scattering function of \citet{Wei01} for a mixture of graphite and silicate 
grains makes no more than a 10\% difference in the derivation of the optical 
constants for $g = 0.65$ and so, for consistency with the literature, we only 
cite the results using the Henyey-Greenstein function. The primary 
uncertainty in our procedure is in the location of the scattering dust and we 
discuss this below. 

The interstellar medium in the direction of Ophiuchus is dominated by the huge
molecular complexes in Ophiuchus (Fig. \ref{f1}) at a distance of $\sim$ 160 
 pc \citep{chen97}. Because of the thickness of the cloud, only foreground stars will 
contribute to the diffuse radiation observed at the Earth. Back-scattering 
from the molecular cloud is an order of magnitude too small to account for the 
observed UV intensity and hence we conclude that the scattering must be 
from two extended sheet-like structures which \citet{Co04} have shown to cover 
the entire region between the galactic longitudes of 290$\degr$ $-$ 10$\degr$ 
and latitudes of -25$\degr$ $-$ +25$\degr$. The nearer of the two sheets is at 
a distance of d $\le$ 60 pc from the Sun and the other, from which most of our 
observed scattering comes, lies between 100 and 150 pc from the Sun, depending 
on the direction. The latter sheet is likely part of the neutral ring 
surrounding the complex of molecular clouds in Ophiuchus discovered by 
\citet{Egger} in the ROSAT all-sky survey. \citet{Co04} has found a column density
of 3.2 $\times\ 10^{19}$ cm$^{-2}$ in the nearer cloud and a much larger column 
density of 3.7 -- 27 $\times\ 10^{20}$ cm$^{-2}$ in the further cloud. Other than 
these two clouds and the Ophiuchus molecular cloud, there is very little material 
out to a distance of at least 200 pc from the Sun \citep{Frisch}.

\begin{figure}
\plotone{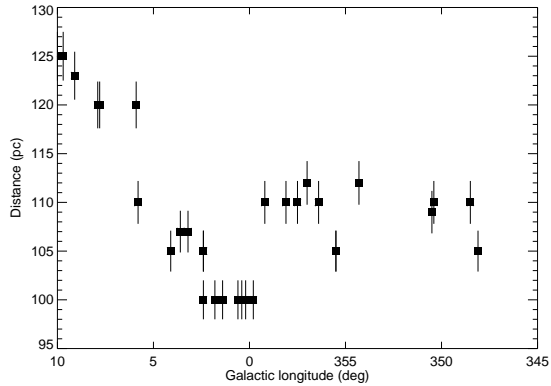}
\caption{Best fit distances to the scattering layer for each of the locations as a function of galactic longitude in degrees.}
\label{f2}
\end{figure}

In our model, we have divided the total H {\small I} from \citet{Dic90} in any 
line of sight into the two foreground clouds, with a constant value of 
3.2 $\times$ 10$^{19}$ cm$^{-2}$ in the nearer cloud and the rest in the 
further cloud. For the Ophiuchus molecular clouds, we have subtracted the 
H {\small I} column density from the total 
hydrogen column density of \citet{Sch98} and distributed this excess material 
at the location of the molecular cloud. We converted the 
H {\small I} column density to a dust scattering cross-section using the 
theoretical values of \citet{Wei01}, implicit in which is the dust to gas ratio 
of \citet{Bohlin}. In practice, the observed UV emission is almost entirely 
from the more distant of the two sheets and so is most sensitive to the exact 
distance of that sheet, or rather to the distance between the sheet and the 
stars dominating the ISRF in this region, and the amount of dust in that cloud.
Because this distance is uncertain, we have used a 3 parameter model in which 
we allow the distance to the dust to vary but fix $a$ and $g$ to a common 
value over all 31 positions. We then use a single scattering model to calculate 
the scattered flux in the UV and independently calculate the thermal emission 
at 100 $\micron$ for the best fit parameters. The best fit distances are plotted
 in Fig. \ref{f2} and show a variation of 100 - 125 pc in this region, 
consistent with the values found by \citet{Co04}. 

\section{RESULTS AND DISCUSSION}

\begin{figure}
\plotone{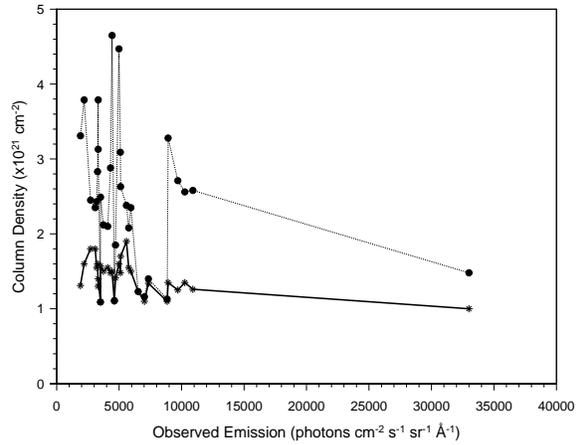}
\caption{Observed UV intensity at each location is plotted against the 
corresponding values of the total, N(H) (dotted line) and neutral, 
N(H {\small I}) (solid line) hydrogen column densities. There is clearly no 
correlation between the FUV intensity and either N(H) or N(H {\small I}).} 
\label{f3}
\end{figure}

\begin{figure}
\plotone{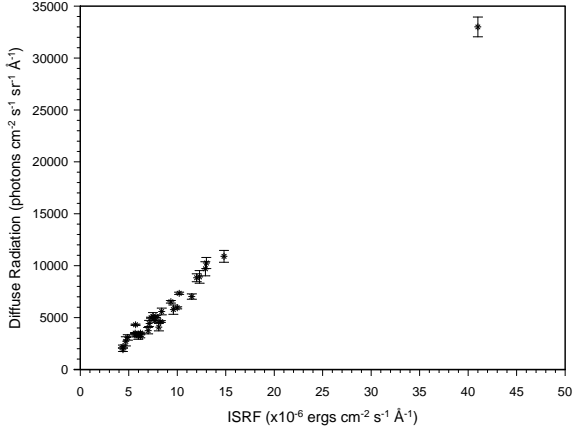}
\caption{Observed UV intensity at each location is plotted against the ISRF. The non-zero intercept is due to the absorption of the diffuse radiation in the intervening ISM.}
\label{f4}
\end{figure}

It is interesting to note that the level of the diffuse emission is not at all 
correlated with the amount of material in the line of sight (Fig. \ref{f3}). 
Instead there is a tight correlation between the level of the ISRF and the 
scattered radiation (Fig. \ref{f4}). This has important implications for the 
study of the diffuse radiation field. It is often claimed that the diffuse 
radiation is correlated with the H {\small I} column density 
\citep[e.g.][]{Bowyer91, Sch}; however, it is becoming increasingly evident 
that the diffuse radiation is dominated by local effects, particularly near 
bright stars \citep[c.f.][]{Mu04}.

\begin{figure}
\plotone{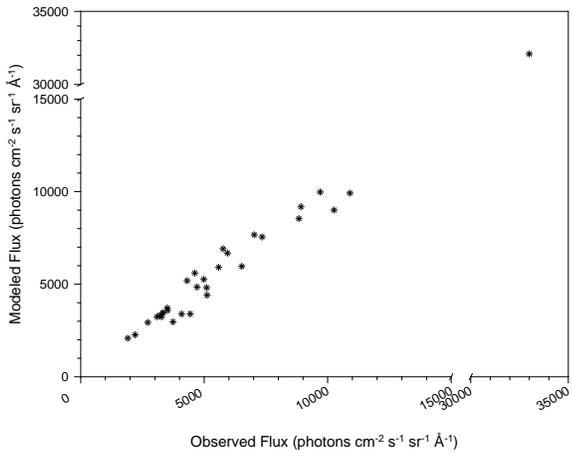}
\caption{Modeled FUV (1100 \AA) intensities corresponding to $a$ = 0.40 
 and $g$ = 0.6 have been plotted against the observed values for each 
location.}
\label{f5}
\end{figure}

\begin{figure}
\plotone{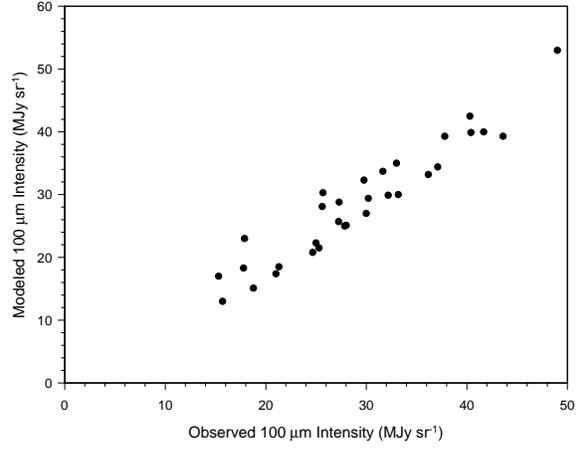}
\caption{Modeled IR intensities corresponding to an albedo of 0.40 and $g$ = 0.6 have been 
plotted against the observed IRAS (100 $\micron$) values for each location.}
\label{f6}
\end{figure}

\begin{figure}
\plotone{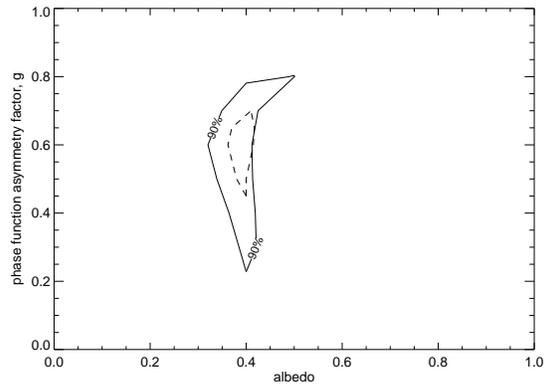}
\caption{90\% confidence contour for all 31 locations (solid contour) is shown. 
The contour corresponds to a limit of 0.40 $\pm$ 0.10 and 0.55 $\pm$ 0.25 on 
the albedo and $g$ respectively. Also plotted is the intersection of the 
individual 90\% confidence contours for each of the locations (dashed line).}
\label{f7}
\end{figure}

Our model predictions match the observations extremely well both in UV 
scattering (Fig. \ref{f5}) and in 100 $\micron$ infrared emission 
(Fig. \ref{f6}). We have derived a 90\% confidence contour 
\citep*[as per][]{lampton} for $a$ and $g$ and this is shown in Fig. \ref{f7}. 
Our 90\% confidence limits on $a$ and $g$ are 0.40 $\pm$ 0.10 and 0.55 $\pm$ 0.25 
respectively and are consistent with the theoretical predictions of \citet{Wei01} 
for average Milky Way dust with R$_{V}$=3.1.

There are very few determinations of the optical parameters of interstellar 
grains at wavelengths shorter than 1200 \AA~\citep{Dr03, Gord04} and most of 
these have come from observations of reflection nebulae. \citet{Witt93} found an 
albedo of 0.42 $\pm$ 0.04 from $\it{Voyager}$ 2 observations of NGC 7023 and 
\citet{Burgh02} found an albedo of 0.30 $\pm$ 0.10 from rocket observations of 
NGC 2023. Both groups claimed that the grains were highly forward scattering with 
$g$ $\sim$ 0.8. \citet{Sh04} found a similar value of 0.40 $\pm$ 0.20 through 
{\it Voyager} observations of the Coalsack Nebula but were not able to constrain $g$.

Although our derived albedo (0.40 $\pm$ 0.10) is consistent with the earlier 
determinations, we find a slightly lower value for $g$ (0.55 $\pm$ 0.25). It 
is possible that conditions are different in reflection nebulae as opposed to 
the diffuse ISM we observe or it may be that, as \citet{Draine} suggests, the 
scattering function is poorly represented by the Henyey-Greenstein function in 
the FUV leading to differences between determinations in different geometries, 
particularly for highly forward scattering grains.

\section{CONCLUSIONS}

We have used {\it Voyager} observations of the diffuse FUV radiation in the 
region of Ophiuchus to investigate the optical constants of the interstellar 
dust grains. We have found that the intense emission in this region arises 
not in the dense molecular cloud which contains most of the matter in this 
direction but rather in a much thinner neutral sheet in front of the cloud. 
In fact, there is no correlation between either 
the 100 $\micron$ IRAS emission or N(H) and the scattered FUV light. Instead 
the FUV emission is tightly correlated with the strength of the local ISRF.
Thus unlike thermal dust emission in the IR, where the dust is optically thin 
and the material along the entire line of sight contributes to the total, 
scattering in the FUV requires much thinner clouds and nearby bright stars. 

In general, one should not expect correlations on global scales between the UV 
and IR or UV and N(H). This is borne out by \citet{Mu04} who find only a mild 
dependence on the 100 $\micron$ for the FUV flux but not by \citet{Sch} who 
claim a correlation between the UV (at 1500 \AA) and N(H) using data from the 
NUVIEWS rocket flight. We are pursuing further investigations with 
data from the Galaxy Evolution Explorer (GALEX) to test these correlations.

We find that the interstellar dust grains are highly forward scattering 
with a $g$ ($= \langle cos \theta \rangle$) of 0.55 $\pm$ 0.25 and 
an albedo of 0.40 $\pm$ 0.10. These results are in general agreement both 
with the theoretical
predictions of \citet{Wei01} and previous observations of reflection nebulae.
As a test of our model, we have independently calculated the 100 \micron ~  
intensities for each of the locations for our best fit $a$ and $g$ values and 
found them to match the IRAS (100 \micron) observations.

\acknowledgments
{We thank an anonymous referee for useful comments which helped to clarify parts of the 
paper. We acknowledge the use of NASA's {\it SkyView} facility 
(http://skyview.gsfc.nasa.gov) located at NASA Goddard Space Flight Center. NVS is 
supported by a RESPOND grant from the Indian Space Reasearch Organization (ISRO). 
PS and JM are supported by the Indian Institute of Astrophysics (IIA), an autonomous 
institution funded by the Department of Science and Technology (DST). RCH is supported 
by Maryland Space Grant Consortium.}
\newpage
\bibliography{ms.bbl}

\clearpage

\clearpage

\clearpage
\begin{deluxetable}{ccccccc}
\tabletypesize{\scriptsize}
\tablecaption{Details of observed locations in Ophiuchus}{\label{t1}}
\tablewidth{0pt}
\tablehead{
\colhead{Location} & \colhead{l} & \colhead{b} & \colhead{Observed Flux$\pm$Error} & \colhead{IRAS(100 $\micron$)$^{a}$} & \colhead{N(H)$^{b}$} \\
& \colhead{(deg)} & \colhead{(deg)} &
\colhead{(ph cm$^{-2}$ s$^{-1}$ sr$^{-1}$ \AA$^{-1}$)} & \colhead{(MJy sr$^{-1}$)} &
\colhead{($\times$ 10$^{21}$ cm$^{-2}$)}
}
\startdata
1&      359.8 &          17.8&           1900$\pm$170 & 43&3.3 \\
2&      0.2   &          22.1&             2200$\pm$140 & 38&3.8 \\
3&      7.9   &          7.8&             2710$\pm$440 & 21&2.5\\
4&      7.8   &          8.4&             3090$\pm$260 & 17&2.4\\
5&      0.6   &          21.4&             3200$\pm$140 & 36&2.4 \\
6&      1.4   &          19.7&             3270$\pm$360  & 30& 2.8 \\
7&      0.4   &          16.8&             3320$\pm$140 & 37&3.8\\
8&      1.8   &          16.2&             3320$\pm$180 & 32&3.1 \\
9&      355.5 &          31&             3510$\pm$50 & 19&1.1\\
10&     5.8   &            21.6&           3510$\pm$75 & 33&2.5\\
11&     2.4   &             19.2&           3740$\pm$310& 26&2.1\\
12&     348.1 &          12.1&           4080$\pm$360  & 40&2.1\\
13&     2.4   &            18.7&           4310$\pm$65 & 32&2.8\\
14&     4.1   &            16.8&           4430$\pm$295  & 30&4.7\\
15&     355.5 &          30.3&           4620$\pm$90 & 16&1.1\\
16&     2.4   &            26.7&           4710$\pm$150  & 27&1.8\\
17&     3.6   &            17&             4990$\pm$50 & 28&4.5\\
18&     3.2   &            18.2&           5100$\pm$90 & 30&3.1\\
19&     5.9   &            14&             5120$\pm$360 & 28&2.6\\
20&     350.4&            9.8&          5580$\pm$325 & 42&2.4 \\
21&     348.5&          12.2&           5760$\pm$445 & 40&2.0\\
22&     359.2&          23.5&           5940$\pm$95 & 33&2.4\\
23&     357.5&          27.5&           6520$\pm$125 & 18&1.2\\
24&     356.4&          29.2&           7020$\pm$260 & 15&1.2\\
25&     358.1&          25.2&           7340$\pm$110 & 26&1.4\\
26&     357&            28.3&           8840$\pm$380 & 18&1.1\\
27&     10.2&           22.4&           8920$\pm$610 & 25&3.3\\
28&     9.1&            23.1&           9700$\pm$680 & 27&2.7\\
29&     9.7&            22.5&           10250$\pm$540& 25&2.6\\
30&     354.3&          13&             10900$\pm$570 & 49&2.6\\
31&     350.5&          24.9&           33000$\pm$955 & 25&1.4\\
\enddata
\label{t1}
$^{a}$ {\citet{wheel}.}\\
$^{b}$ {Column densities from \citet{Sch98}.}
\end{deluxetable}

\clearpage
\begin{deluxetable}{cccccccc}
\tabletypesize{\scriptsize}
\tablecaption{Properties of contributing stars}{\label{t1}}
\tablewidth{0pt}
\tablehead{
\colhead{Star} & \colhead{l} & \colhead{b} & \colhead{Spectral Type} & \colhead{Temperature} & \colhead{log (g)} & \colhead{log (z)} \\
& \colhead{(deg)} & \colhead{(deg)} & & \colhead{(K)} &  & 
}
\startdata
$\zeta$ Oph   &    6.28   &    23.59 &   O9V    &   35000      & 3.94 & 0 \\
$\chi$ Oph    &    357.93 &    20.68 &   B2Vne  &   22000      & 3.94 & 0 \\
$\nu$ Sco A   &    354.61 &    22.70 &   B2IV   &   22000      & 3.94 & 0 \\
$\omega$ Sco  &    352.75 &    22.77 &   B1V    &   25600      & 3.94 & 0 \\
$\delta$ Sco  &    350.10 &    22.49 &   B0.2IV &   30000      & 3.94 & 0 \\
$\pi$ Sco     &    347.22 &    20.23 &   B1V    &   25600      & 3.94 & 0 \\
$\tau$ Sco    &    351.54 &    12.81 &   B0V    &   30000      & 3.94 & 0 \\
$\lambda$ Lib &    350.72 &    25.38 &   B3V    &   19000      & 3.94 & 0 \\
\enddata
\label{t2}
\end{deluxetable}

\end{document}